\documentstyle[prl,twocolumn,aps]{revtex}
\begin{document}
%\eqnobysec   %JPB
%\jl{2}    %JPB
%\twocolumn[
\draft    %JPB

\title{Quantum effects on curve crossing in 
a Bose-Einstein condensate}

\author{V. A. Yurovsky$^{1,2,3}$, A. Ben-Reuven$^{2}$, and P. S.
 Julienne$^{3}$}

\address{$^{1}$ITAMP, Harvard-Smithsonian Center for
Astrophysics, 60 Garden Street, Cambridge MA 02138}

\address{$^{2}$School of Chemistry, Tel Aviv University, 69978 Tel
 Aviv,
Israel}

\address{$^{3}$Atomic Physics Division, Stop 8423, National
Institute of Standards and Technology, Gaithersburg, MD 20889}

\date{\today}
\maketitle    %JPB
%\widetext

\begin{abstract} Formation of atomic pairs by the dissociation of
a molecular condensate or by inelastic collisions in an atomic
condensate due to a time-dependent curve crossing 
process is studied beyond the mean-field 
approximation. The number of atoms formed by the spontaneous process
is described by a Landau-Zener formula 
multiplied by an exponential amplification factor due to quantum
many-body effects. The atomic pairs are 
formed in an entangled (squeezed) state. The rate of stimulated 
processes depends on the relative phase of the two fields.
\end{abstract}
%]
\narrowtext

\paragraph*{Introduction.---}
The concept of curve crossing was introduced in the pioneering works
of Landau \cite{L32} and Zener \cite{Z32} concerning transitions
between two linear potentials. Curve-crossing
theory has many applications in various fields of physics (see, e.g.,
Ref.\ \cite{YB01} and references therein), usually describing
transitions between states whose energies cross as a function
of the time (a non-stationary crossing) or some coordinate (a
stationary crossing).

Recent developments in the physics of Bose-Einstein condensates (BEC)
(see Refs.\ \cite{PW98,DGPS99} and references therein) demand taking 
account of many-body effects in a curve crossing. These effects, which
cannot be described by the usual mean-field method, are treated here
by a second-quantized formalism. Hereafter we
call a process involving such effects a ``many-body crossing'', 
as distinct from a ``single-body crossing'' in which a colliding pair
of atoms is treated as a single body (a quasimolecule).

A major many-body effect on curve-crossing in BEC is Bose
enhancement. In the photoassociation of atomic BEC this effect leads to
Bose-stimulated Raman adiabatic passage (see Ref.\ \cite{MKJ00} and
references therein). An expression for the transition probability as a
function of the rate of energy variation was obtained in
\cite{YBJW99,YBJW00} by treating the process as a curve
crossing event in a nonlinear system. The nonlinearity may lead to a
deviation of the transition probability from that predicted by the
well-known Landau-Zener (LZ) formula.

Although the association of atoms into molecules
has been described within a mean-field approach (using coupled
Gross-Pitaevskii equations), the process of dissociation of
molecules into  the atomic vacuum 
\begin{equation}
A_2\rightarrow A+A \label{A2diss}
\end{equation}
cannot be described by
this approach, since the coupling is proportional to the
atomic field, the initial value of which is zero. Therefore, at least
initially, the dissociation must be a spontaneous process, which is
neglected in the mean-field approach.

The time-independent dissociation of a molecular BEC was considered in 
Ref.\ \cite{VYA01} by using a numerical   
solution of the full quantum equations.
It is shown there that the initial stage of the
dissociation can be approximately described with good accuracy
by analytical expressions obtained by treating
the molecular field as a mean field, while the atomic 
field is treated as a second-quantized field. This method is similar to 
the parametric approximation used in quantum optics (see Ref.\ \cite{PA}).
This approximation is valid as long as the population of the molecular 
state is large and the effect of its depletion can be neglected.

In the present paper we generalize the parametric approximation to the
case of time-dependent condensate energies. The energies can be varied by 
applying an external time-dependent magnetic field or by changing
the frequency of laser fields coupling the states. In general, the 
dissociation of a molecular condensate requires taking into account
the coupling to multiple atomic modes, in a manner analogous to methods of multistate curve crossing theory 
\cite{YB01,ChildMCT}.
Here we concentrate on the essential physics by considering only a single
atomic mode, showing how quantum effects lead to a substantial initial 
enhancement of the dissociation and to the formation of  atoms 
in squeezed states that are entangled  (see also \cite{VA01,AV01,PM01}).

Similar quantum effects can also play an important role in the
association of atomic condensates, where the molecular state is
coupled to excited atomic trap states (which are initially not
populated).  Transitions to these states can substantially affect the
condensate loss rate \cite{YBJW00,MJT99} and therefore should also be
treated by a multimode model using a quantum approach similar to the
one described below.

The effects of Bose enhancement and the formation of entangled
atoms should also appear in an atomic BEC, if inelastic
collisions of the type
\begin{equation}
A(0)+A(0)\rightarrow A+A , \label{Acoll} 
\end{equation}
where $A(0)$ and $A$
denote different internal states of the atoms, change the states of 
both colliding atoms. Such a process might be 
realized by applying an electromagnetic field, or two fields
(as in a Raman process), in resonance with the two-atom transition, 
provided that the concurrent one-atom transition ($A(0)\rightarrow A$)
is off resonance with the applied fields. The time variation of
the energy can again be due to the Zeeman effect or to 
time-dependent electromagnetic field frequencies.
An experimental observation of Bose enhancement and entanglement 
may be simpler in an atomic BEC than in the molecular one.

\paragraph*{General crossing.---} \label{General}
Consider a system of coupled atomic $A$ and molecular $A_2$ fields,
associated with the annihilation operators $\hat{\psi }$  and $\hat{\psi
 }_{m}$, respectively. In the case of one atomic and one molecular mode 
the system can be described by the Hamiltonian
\begin{equation}
\hat{H}=\epsilon \left( t\right) \hat{\psi }^{\dag }\hat{\psi }
+g_{\text{am}}\hat{\psi }^{\dag }_{m}\hat{\psi }\hat{\psi }+g^{
*}_{\text{am}}\hat{\psi }^{\dag }\hat{\psi }^{\dag }\hat{\psi }_{m} ,
\end{equation}
%V
where $\epsilon \left( t\right) $ is a time-dependent energy of the 
atomic state, while
the energy of the molecular state is chosen as the zero energy. The
parameter $g_{\text{am}}$ describes the coupling of the atomic and
molecular fields due to hyperfine interaction or laser fields.

The parametric approximation used in Ref.\ \cite{VYA01} for
the case of constant $\epsilon$ can be generalized to the case of a
time-dependent
$\epsilon $, by retaining the assumption of a large molecular population.
We can then replace the operator
$\hat{\psi }_{m}$ by its expectation value $\langle \hat{\psi
 }_{m}\rangle $, and obtain the
equation of motion for the atomic field operator $\hat{\psi }$:
\begin{equation}
i \dot{\hat{\psi }} = \epsilon \left( t\right) \hat{\psi } + 2g^{
*}\hat{\psi }^{\dag } \label{pside1}
\end{equation}
(using units in which $\hbar=1$), where 
\begin{equation}
g=g_{\text{am}}\langle \hat{\psi }_{m}\rangle ^*  . \label{g}
\end{equation}

Let us also consider two atomic states, $A(0)$ and $A$, associated
with the field operators $\hat{\psi}_0$ and $\hat{\psi}$, respectively.
Taking into account one mode for each of the states, one can describe 
the system by the following Hamiltonian
\begin{equation}
\hat{H}=\epsilon \left( t\right) \hat{\psi }^{\dag }\hat{\psi }
+g_{\text{aa}}\hat{\psi }^{\dag }_0\hat{\psi }^{\dag }_0\hat{\psi }\hat{\psi }
+g^*_{\text{aa}}\hat{\psi }^{\dag }\hat{\psi }^{\dag }\hat{\psi }_0\hat{\psi }_0 ,
\end{equation}
where $\epsilon \left( t\right)$ is the time-dependent energy of state $A$, 
while the energy of state $A(0)$ is chosen as the zero energy. Whenever both 
wavevectors $k_0$ and $k$, of the entrance and exit reaction channels 
(\ref{Acoll}), are in the near-threshold region, the 
inelastic scattering amplitude should have the form $T=\sqrt{k_0 k}b$,
where $b$ is an energy-independent length 
(see Ref.\ \cite{TS}). The parameter $g_{\text{aa}}$ can then be expressed as 
$g_{\text{aa}}=4\pi b/m$. The parametric approximation can be
applied here if the state $A(0)$ has a large population, replacing the product
$\hat{\psi }_0\hat{\psi }_0$ by its expectation value
$\langle\hat{\psi }_0\hat{\psi }_0\rangle$. If the atoms are initially in a
condensate (coherent) state, then 
$\langle\hat{\psi }_0\hat{\psi }_0\rangle=\langle\hat{\psi }_0\rangle^2$.
Denoting
\begin{equation}
g=g_{\text{aa}}\langle\hat{\psi }_0\hat{\psi }_0 \rangle ^*  , \label{ga}
\end{equation}
one obtains again the same equation of motion (\ref{pside1}) for the
field operator $\hat{\psi }$.

The set of two first-order equations [Eq.\ (\ref{pside1}) and its
hermitian conjugate] can be transformed to a single second-order equation
\begin{equation}
\ddot{\hat{\psi }} +\left( \epsilon ^{2}\left( t\right) -4|g|^{2}
+i\dot{\epsilon }\left( t\right) \right) \hat{\psi } =0 .
 \label{pside2}
\end{equation}
Although it is an operator equation, it is linear and its
solution can be expressed as
\begin{equation}
\hat{\psi }\left( t\right) =\psi _{c}\left( t\right) \hat{\psi }\left
( t_{0}\right) +\psi _{s}\left( t\right) \hat{\psi }^{\dag }\left(
 t_{0}\right)  , \label{psisol}
\end{equation}
where $\psi _{c,s}\left( t\right) $ are two independent $c$-number
 solutions of Eq.\
(\ref{pside2}).
Equations (\ref{pside1}) and (\ref{psisol}) at $t=t_{0}$  give 
the initial conditions for $\psi _{c,s}\left( t_{0}\right) $
and their derivatives:
\begin{eqnarray}
&&\psi _{c}\left( t_{0}\right) =1,\qquad \psi _{s}\left( t_{0}\right)
 =0,  \label{psit0}
\\
&&\dot{\psi }_{c}\left( t_{0}\right) =-i\epsilon \left( t_{0}\right)
 ,\qquad \dot{\psi }_{s}\left( t_{0}\right) =-2i g^{*} .
 \label{dpsit0}
\end{eqnarray}

The solutions $\psi _{c}\left( t\right) $ and $\psi _{s}\left( t\right) $ 
can be expressed in terms of a pair of conventionally selected independent 
solutions of Eq.\
(\ref{pside2}) $\psi _{1}\left( t\right) $ and $\psi _{2}\left( t\right) $, 
treated as ``standard solutions'', in the form 
\begin{eqnarray}
\psi _{c}\left( t\right) =&& {i\over W\{\psi _{1},\psi
 _{2}\}}\Bigl\lbrack \left( \epsilon \left( t_{0}\right) \psi
 _{2}\left( t_{0}\right) -i\dot{\psi }_{2}\left( t_{0}\right) \right)
 \psi _{1}\left( t\right)  \nonumber
\\
&&-\left( \epsilon \left( t_{0}\right) \psi _{1}\left( t_{0}\right)
-i\dot{\psi }_{1}\left( t_{0}\right) \right) \psi _{2}\left( t\right)
 \Bigr\rbrack
,\\
\psi _{s}\left( t\right) =&&{2i g^*\over W\{\psi _{1},\psi
 _{2}\}}\left\lbrack \psi _{2}\left( t_{0}\right) \psi _{1}\left(
 t\right) -\psi _{1}\left( t_{0}\right) \psi _{2}\left( t\right)
 \right\rbrack  ,\label{psis}
\end{eqnarray}
where
$W\{\psi _{1},\psi _{2}\}=\psi _{1}\dot{\psi }_{2}-\psi _{2}\dot{\psi}_{1}$
is the Wronskian of the standard solutions. In the special case of a 
time-independent $\epsilon $ \cite{VYA01}, $\psi _{s}\left( t\right)$ and 
$\psi _{c}\left( t\right)$ can be  expressed in terms of
sine and cosine functions (circular or hyperbolic, depending on 
the ratio of $g$ to $\epsilon $).

The use of Eq.\ (\ref{psisol}) allows us to express the expectation 
value of any function of $\hat{\psi }^{\dag }$  and $\hat{\psi }$
in terms of the expectation value at a particular time $t=t_{0}$. 
Thus, the number of atoms is a sum of two terms
$\langle \hat{\psi }^{\dag }\left( t\right) \hat{\psi }\left( t\right)
 \rangle =n_{sp}\left( t\right) +n_{st}\left( t\right)$  ,
where the first term
\begin{equation}
n_{sp}\left( t\right) =|\psi _{s}\left( t\right) |^{2} \label{nsp}
\end{equation}
corresponds to spontaneous transitions into the vacuum of $\hat{\psi }$,
and the second term
\begin{eqnarray}
n_{st}\left( t\right) =&&\left( |\psi _{c}\left( t\right) |^{2}+|\psi
 _{s}\left( t\right) |^{2}\right) \langle \hat{\psi }^{\dag }\left(
 t_{0}\right) \hat{\psi }\left( t_{0}\right) \rangle  \nonumber
\\
&&+2\text{Re}\left( \psi ^{*}_{s}\left( t\right) \psi _{c}\left(
 t\right) \langle \hat{\psi }\left( t_{0}\right) \hat{\psi }\left(
 t_{0}\right) \rangle \right)  . \label{nst}
\end{eqnarray}
depicts stimulated transitions taking place when the final atomic state is
initially populated. The first and second terms on the right-hand side 
of Eq.\ (\ref{nst}) correspond to a noncoherent and a coherent initial 
occupation, respectively.
Consider also the
expectation value of a generalized coordinate operator
\begin{equation}
\hat{Q}(t)=\hat{\psi }(t)e^{i\theta }+\hat{\psi }^{\dag }(t)e^{-i\theta
 } \label{Q}
\end{equation}
frequently used in the theory of
squeezed states (see, e. g., \cite{MW95}). We present here only an 
expression for the spontaneous process, 
in which $\langle \hat{\psi }^{\dag }\left( t_{0}\right) \hat{\psi }\left
( t_{0}\right) \rangle =\langle \hat{\psi }\left( t_{0}\right)
 \hat{\psi }\left( t_{0}\right) \rangle =0$:
\begin{equation}
\langle \hat{Q}^{2}(t)\rangle _{sp}=|\psi _c(t)e^{i\theta}+
\psi_s^*(t)e^{-i\theta} |^{2} . \label{Qsp}
\end{equation}

\paragraph*{Linear crossing.---} \label{Linear}
Single-body curve-crossing problems are commonly solved by using the
LZ formula, which is an exact solution of the linear non-stationary
problem.  Let us consider a generalization to our case of a many-body
crossing, choosing $\epsilon \left( t\right )$ as the linear function
$\epsilon \left( t\right) =\beta t$.
The second-order equation (\ref{pside2}) attains in this case the form 
of a parabolic cylinder equation (see Ref.\ \cite{Abramovitz})
\begin{equation}
\ddot{\hat{\psi }} +\left( \beta ^{2}t^{2}-4|g|^{2}+i\beta \right)
 \hat{\psi } =0 . \label{pside2l}
\end{equation}
This equation differs from the corresponding c-number equation for 
a single-body crossing (see Ref.\ \cite{ChildMCT}) by the sign before 
$4|g|^2$ due to the effects of secondary quantization.
An appropriate choice of a pair of standard solutions is then
\begin{eqnarray}
&&\psi _{1}\left( t\right) =U\left( {1\over 2} +i\lambda ,-e^{-i\pi
/4}\tau \right)  ,\nonumber
\\
\label{psi12}
\\
&&\psi _{2}\left( t\right) ={1\over g}\sqrt{\beta /2} e^{
-i\pi /4}U\left( -{1\over 2} -i\lambda ,-e^{i\pi /4}\tau \right)  ,
 \nonumber
\end{eqnarray}
where $U\left( a,x\right) $ is the parabolic cylinder function, 
defined by Eq.\ (19.3.1) of Ref.\ \cite{Abramovitz}. Here  
$\tau =\sqrt{2\beta } t$, and
\begin{equation}
\lambda =2|g|^{2}/\beta . \label{lambda}
\end{equation}
The parameter  $\lambda $ is nothing else but the LZ exponent, since
the coupling of states, according to Eq.\ (\ref{pside1}), is $2g$ and
the slope of the two-atom energy is $2\beta $.

The solutions (\ref{psi12}) have the Wronskian
$W\{\psi _{1},\psi _{2}\}=i(\beta/ g) \exp(-\pi \lambda /2)$
and in addition obey the conditions
\begin{equation}
i\dot{\psi }_{i}-\beta t\psi _{i}=2g^{*}\psi ^{*}_{3-i}\qquad \left(
 i=1,2\right) ,
\end{equation}
which allow us to express $\psi _{c}\left( t\right) $ as
\begin{equation}
\psi _{c}\left( t\right) =-2i g^* 
{ \psi ^{*}_{1}\left( t_{0}\right) \psi _{1}\left(
 t\right) -\psi ^{*}_{2}\left( t_{0}\right) \psi _{2}\left( t\right)
\over W\{\psi _{1},\psi _{2}\}}  . \label{psic}
\end{equation}

Using a common procedure of curve crossing theories, 
we can start by considering transitions during a finite time interval 
$\left\lbrack t_{0},t\right\rbrack $, with $t_{0}<0<t$, 
assuming sufficiently large values of 
$|t_{0}|$ and $t$ (obeying the validity criteria presented 
below).
Asymptotic expansions of the parabolic cylinder functions
\cite{Abramovitz} give us at $t_{0}\rightarrow -\infty $
\begin{eqnarray}
\psi _{1}\left( t_{0}\right) \sim {1\over |\tau _{0}|}\exp\left(
-{\pi \over 4}\lambda +i{\pi \over 4}+i S\left( |\tau _{0}
|\right) \right)  ,\nonumber
\\
\label{ast0}
\\
\psi _{2}\left( t_{0}\right) \sim {1\over g}\sqrt{{\beta
 \over 2}} \exp\left( -{\pi \over 4}\lambda -i{\pi \over 4}
-iS\left( |\tau _{0}|\right) \right)  ,\nonumber
\end{eqnarray}
where  $\tau _{0}=\sqrt{2\beta } t_{0}$  and $S\left( \tau \right)
 =\tau ^{2}/4-\lambda \ln \tau $. The asymptotic expansions at
$t\rightarrow \infty $, obtained by using the relation (19.4.6) in
 \cite{Abramovitz}, are given by
\begin{eqnarray}
&&\psi _{1}\left( t\right) \sim {1\over \tau }\exp\left( {3\pi \over
 4}\lambda -i{5\pi \over 4}+i S\left( \tau \right) \right)
 \nonumber
\\
&&+ {1\over |g|}\sqrt{\beta \sinh\left( \pi \lambda \right) }
 \exp\left( {\pi \over 4}\lambda  - i{\pi \over 2}-i S\left(
 \tau \right) -i \arg\Gamma \left( i\lambda \right) \right)  ,\nonumber
\\
\label{ast}
\\
&&\psi _{2}\left( t\right) \sim {1\over g}\sqrt{\beta /2}
 \exp\left( {3\pi \over 4}\lambda  -i{\pi \over 4}-i S\left(
 \tau \right) \right)  \nonumber
\\
&&+ {g^*\over |g|\tau }\sqrt{2\sinh\left( \pi \lambda \right) }
 \exp\left( {\pi \over 4}\lambda  -i{\pi \over 2}+i S\left(
 \tau \right) +i \arg\Gamma \left( i\lambda \right) \right)  .\nonumber
\end{eqnarray}

%\paragraph{Results and discussion} \label{ResDisc}

\paragraph*{Spontaneous process.---}
Substituting the asymptotic expansions, Eqs.\ (\ref{ast0}) and
(\ref{ast}), at large $|t_0|$ and $t$ into Eqs.\ (\ref{psis}) and
(\ref{nsp}), one can obtain the number of atoms formed by the 
spontaneous process
\begin{eqnarray}
n_{sp}\left( t\right) \sim \left( e^{2\pi \lambda }-1\right)
+&&2e^{\pi \lambda }\sqrt{\lambda \left(e^{2\pi \lambda }-1\right)}
 \nonumber
\\
&&\times \left\lbrack {\cos \chi(\tau) \over \tau }
-{\cos \chi(|\tau _0|)\over \tau _0}\right\rbrack , \label{spd}
\end{eqnarray}
where
$\chi \left( \tau \right) =\tau^2/ 2-2 \lambda  \ln \tau
  +\arg \Gamma \left( i\lambda \right)$.

Equation (\ref{spd}) contains two terms of an asymptotic expansion in
inverse powers of $\tau $ and $\tau _{0}$.  The asymptotic approach is
applicable if the second term in Eq.\ (\ref{spd}) is small compared to
the first one, or
\begin{equation}
t , |t_0|\gg |g/\beta|  ,
\end{equation}
in agreement with estimates of the 
transition region for two-state single-body crossings 
(see Ref.\ \cite{ChildMCT}).

Let us recall that all the analysis above is based on the assumption
that the depletion of the initial (atomic or molecular) condensate 
state is negligibly small, applicable when
$n_{sp}\left( t\right) \ll N$, or
\begin{equation}
\lambda < {1\over 2\pi }\ln N ,
\end{equation}
where $N$ is the initial number of atoms. Since $N\gg 1$ (in current
experiments an atomic BEC contains $10^{4}$ to $10^{6}$ atoms) the 
present approach is applicable even for large values of $\lambda $ 
corresponding to adiabatic transitions.

The leading term of Eq.\ (\ref{spd}) can be written as
\begin{equation}
n_{sp}\approx \left( 1-e^{-2\pi \lambda }\right) e^{2\pi \lambda } .{
 } ^{\label{spd0}}
\end{equation}
The first factor here is the familiar LZ
probability, while the second one describes an amplification of the
spontaneous process. This amplification is reminiscent of lasing, 
as the exponent can be interpreted as a product of a characteristic 
crossing time ($g/\beta $) and
an ``amplification coefficient'' $g$.

The LZ exponent $\lambda $ is itself enhanced, being proportional to $N$ 
[see Eqs.\ (\ref{g})and (\ref{lambda})].  This is also true for the 
{\it association} process, but unlike that process, the Bose enhancement 
introduced here by the amplification factor is unique to the dissociation 
process.  As an example, consider the Na Feshbach
resonances studied in Ref.~\cite{YBJW00}. The parameter $\lambda$ there
can be of the order of unity, depending on the ramp speed, in which case
one obtains an amplification factor of more than a 100.

A distinction between many-body and single-body crossings appears 
also in the interpretation of the crossing as scattering on a
parabolic potential barrier. Indeed, on replacing the time by a
coordinate, Eq.\ (\ref{pside2}) is transformed into a corresponding 
one-dimensional stationary Schr\"odinger equation. In the single-body 
case the energy lies above the 
barrier (see Ref.\ \cite{ChildMCT}). However, in the many-body case, 
as one can see from 
Eq.\ (\ref{pside2}), the energy lies below the the barrier,
and the process involves barrier penetration. 
This leads to the replacement of the negative
exponents by positive ones, and yields the first term of Eq.\
(\ref{spd}) in place of the LZ formula.

Using Eqs.\ (\ref{psis}), (\ref{psic}), (\ref{ast0}), and (\ref{ast}),
we can write Eq.\ (\ref{Qsp}) as
\begin{eqnarray}
\langle \hat{Q}^{2}\rangle _{sp}&&\sim \Bigl | \sqrt{e^{2\pi
 \lambda }-1} \nonumber 
\\
&&+\exp\left( \pi \lambda -i\chi \left( \tau \right)
-i \arg g+2i\theta -i\pi /2\right) \Bigr | ^{2} .
\end{eqnarray}
Substituting one of the two orthogonal phase angles 
$\theta =\theta _{\pm }={1\over 2}(\chi( \tau)+\arg g)\pm 
{\pi \over 4}$ in Eq.\ (\ref{Q}) we obtain
\begin{equation}
\langle \hat{Q}^{2}_{\pm }\rangle _{sp}\sim  \left |
 \sqrt{e^{2\pi \lambda }-1}\pm e^{\pi \lambda }\right |
 ^{2}\mathrel{\mathop\approx _{\lambda\gg 1}} \cases{4e^{2\pi \lambda
 }\cr {1\over 4}e^{-2\pi \lambda }}.
\end{equation}
Therefore, the spontaneous dissociation forms atoms in a squeezed
state (similar to a squeezed state of light \cite{MW95}) in which
 $\hat{Q}_{+}$
has an increased uncertainty and $\hat{Q}_{-}$ has a reduced 
uncertainty. A similar effect concerning the time-independent case
has been described in Ref.\ \cite{VYA01}.

\paragraph*{Stimulated process.---}
Consider now the transition stimulated by a finite initial
occupation of the final atomic state. Starting from a coherent state
$|\alpha \rangle $ of the operator $\hat{\psi }$, such that
$\hat{\psi }\left( t_{0}\right) |\alpha \rangle =\alpha |\alpha\rangle$,
and using Eqs.\ (\ref{nst}), (\ref{ast0}), and (\ref{ast}),
one obtains
\begin{eqnarray}
n_{st}&&\sim |\alpha |^{2} \Bigl | \sqrt{e^{2\pi \lambda }-1}
 \nonumber
\\
&&+\exp\left( \pi \lambda +i\chi \left( \tau _{0}\right) 
+i\arg g+2 i \arg \alpha  +i\pi /2\right) \Bigr | ^{2} .
\end{eqnarray}
It follows from this expression that the final atomic population is
dependent on the phase of $\alpha $, lying in a range bounded by the
maximal and minimal values
\begin{equation}
n_{\pm }=|\alpha |^{2}\left | \sqrt{e^{2\pi \lambda }-1}\pm
 e^{\pi \lambda }\right | ^{2}\mathrel{\mathop\approx _{\lambda\gg
 1}} |\alpha |^{2}\cases{4e^{2\pi \lambda }\cr {1\over 4}e^{-2\pi
 \lambda }}
\end{equation}
at $\arg \alpha =-{1\over 2}(\chi(|\tau _{0}|)+\arg g)\mp {\pi \over
 4}$, respectively. Since $\arg g$ contains the phase
 of the molecular field $\langle \hat{\psi }_{m}\rangle $
 [see Eq.\ (\ref{g})] or the atomic field $\langle \hat{\psi }_{0}\rangle $
 [see Eq.\ (\ref{ga})], the stimulated dissociation is indeed dependent on the
phase difference between the two fields ($\alpha $ and 
$\langle \hat{\psi }_{m}\rangle $ or $\langle \hat{\psi }_{0}\rangle $).
This effect is reminiscent of the amplification or absorption of 
coherent light in laser-active media, and its dependence on the phase 
of the light.

\paragraph*{Conclusions.---}
The parametric approximation allows us to obtain an analytical 
expression (\ref{spd0}) for the number of atoms spontaneously formed by 
the crossing of linearly time-dependent energies of the atomic and the 
molecular states or two atomic states. Many-body effects 
lead to an enhancement factor 
in the LZ exponent, as well as to the multiplication of the LZ probability by 
the amplification factor in Eq.\ (\ref{spd0}) describing a positive-exponential 
enhancement of the spontaneous process. The atoms are formed in 
a squeezed state.  The number of atoms formed by the stimulated process
depends on the relative initial phase of 
the two fields. Similar quantum effects should also apply 
in other cases in which particles are created in pairs. An experimental
realization of the processes we discuss may allow production of an 
entangled atomic gas.

%\paragraph*{Acknowledgments.---}
The authors are grateful to P. D. Drummond and K. V. Kheruntsyan for helpful
discussions. This work was partially supported by the NSF 
through a grant for the ITAMP at Harvard University and the SAO.


\begin{references}
\bibitem{L32}L. D. Landau, Phys.\ Z.\ Sowjetunion {\bf 2}, 46 (1932).
\bibitem{Z32}C. Zener, Proc.\ R.\ Soc.\ London Ser.\ A {\bf 137},
696 (1932).
\bibitem{YB01} V. A. Yurovsky and A. Ben-Reuven, Phys. Rev A
{\bf 63}, 043404 (2001).
\bibitem{PW98} A. S. Parkins and D. F. Walls, Phys. Rep. {\bf 303}, 1
(1998).
\bibitem{DGPS99} F. Dalfovo, S. Giorgini, L. P. Pitaevskii, and S.
Stringari, Rev. Mod. Phys. {\bf 71}, 463 (1999).
\bibitem{MKJ00} M. Mackie, R. Kowalski, and J. Javanainen, Phys. Rev.
Lett. {\bf 84}, 3803 (2000).
\bibitem{YBJW99} V. A. Yurovsky, A. Ben-Reuven, P. S. Julienne and C. J.
Williams, Phys. Rev. A {\bf 60}, R765 (1999).
\bibitem{YBJW00} V. A. Yurovsky, A. Ben-Reuven, P. S. Julienne and C. J.
Williams, Phys. Rev. A {\bf 62}, 043605 (2000).
\bibitem{VYA01}A. Vardi, V. A. Yurovsky, and J. R. Anglin, Phys. Rev. A
(submitted).
\bibitem{PA}M. O. Scully and M. S. Zubairy, {\it Quantum Optics} 
(University Press, Cambridge, 1997). 
\bibitem{ChildMCT}M. S. Child, {\it Molecular Collision Theory}
(Academic Press, London and New York, 1974).
\bibitem{VA01}
A. Vardi and J. R. Anglin, Phys. Rev. Lett. {\bf 86}, 568 (2001).
\bibitem{AV01}
J. R. Anglin and A. Vardi, Phys. Rev. A, {\bf 64}, 013605 (2001).
\bibitem{PM01} U. V. Poulsen and K. Molmer, Phys. Rev. A {\bf 63},
 023604 (2001). 
\bibitem{MJT99} F. H. Mies, E. Tiesinga, and P. S. Julienne,
Phys.Rev. A {\bf 61}, 022721 (2000).
\bibitem{TS}P. S. Julienne and F. H. Mies, J. Opt. Soc. Am. B {\bf 6},
2257 (1989); H. R. Sadeghpour {\it et al.}, J. Phys. B {\bf 33},
R93 (2000).
\bibitem{MW95}L. Mandel and E. Wolf, {\it Optical Coherence and
Quantum Optics} (University Press, Cambridge, 1995).
\bibitem{Abramovitz} {\it Handbook of Mathematical Functions}, edited
 by M. Abramowitz and I. E. Stegun (NBS, Washington, 1964).
\end{references}
\end{document}